\documentstyle[aps,12pt]{revtex}

\begin{document}
\title{B. B. G. K. Y. Hierarchy Methods for Sums of Lyapunov Exponents for Dilute
Gases}
\author{J. R. Dorfman$^{1}$, Arnulf Latz$^{2}$, and Henk van Beijeren$^{3}$}
\address{${1}$) Institute for Physical Science and Technology and\\
Department of Physics, \\
University of Maryland, College Park, MD,\\
20742, USA \\
${2}$) Fachbereich Physik (18), Institut f\"ur Physik \\
Johannes Gutenberg\\
Universit\"at Mainz, \\
D-55099 Mainz, Germany \\
${3}$) Institute for\\
Theoretical Physics, University of Utrecht, \\
Postbus 80006, Utrecht,\\
3508 TA,\\
The Netherlands}
\date{\today}
\maketitle

\begin{abstract}
We consider a general method for computing the sum of positive Lyapunov
exponents for moderately dense gases. This method is based upon hierarchy
techniques used previously to derive the generalized Boltzmann equation for
the time dependent spatial and velocity distribution functions for such
systems. We extend the variables in the generalized Boltzmann equation to
include a new set of quantities that describe the separation of trajectories
in phase space needed for a calculation of the Lyapunov exponents. The
method described here is especially suitable for calculating the sum of all
of the positive Lyapunov exponents for the system, and may be applied to
equilibrium as well as non-equilibrium situations. For low densities we
obtain an extended Boltzmann equation, from which, under a simplifying
approximation, we recover the sum of positive Lyapunov exponents for hard
disk and hard sphere systems, obtained before by a simpler method. In
addition we indicate how to improve these results by avoiding the
simplifying approximation. The restriction to 
hard sphere systems in $d$-dimensions is made to keep the somewhat
complicated formalism as clear as possible, but the method can be easily
generalized to apply to gases of particles that interact with strong short
range forces.
\end{abstract}


\section{Introduction}

The ergodic and mixing properties of classical $d$-dimensional hard sphere
systems has been the subject of a great deal of work over the past several
decades. Stimulated by the progress made by Sinai and co-workers toward
proving the ergodicity of billiard type systems in the 1960's and onward\cite
{sinai}, others have extended these results in a number of directions,
including a recent proof of the ergodic behavior of some systems of hard
spheres in $2$ and $3$ dimensions by Szasz and Simanyi\cite{szsim}.
Moreover, numerical studies of the chaotic behavior of hard sphere systems
have recently been carried out by Dellago, Posch and Hoover who computed the
Lyapunov spectrum, and the Kolmogorov-Sinai (KS) entropy, $h_{KS}$ in the
case that the spheres are in equilibrium or in non-equilibrium steady states%
\cite{delpoho}. For closed systems, according to Pesin's theorem, the KS
entropy is equal to the sum of the positive Lyapunov exponents\cite
{ru-eck,pesin,kathas}.

For a simple model system, the Lorentz gas at low densities, it has been
possible to calculate analytically the Lyapunov spectrum and to show that
the results so obtained are in good agreement with computer results \cite
{lorpaps}. This model considers the motion of a point particle moving in a
random array of fixed scatterers, which are usually taken to be hard spheres
in $d$ dimensions. 
It was shown that standard kinetic theory methods for treating the Lorentz
gas could be extended to include a systematic calculation of the Lyapunov
spectrum. Methods used included simple free path analyses as well as an
extension of the Lorentz-Boltzmann transport equation to include new
variables which describe the separation of trajectories in the phase space
of the moving particle. Thus the question arises as to whether it is
possible to extend the Boltzmann transport equation and its generalization
to higher densities, as well as other related methods in kinetic theory, to
compute properties of the Lyapunov spectrum of a gas where all of the
particles move. Some results in this direction have recently been obtained.

A theoretical treatment of the Lyapunov spectrum and $h_{KS}$ has recently
been undertaken for hard disks or spheres at low density, $n\sigma^{d} << 1$%
, where $n$ is the number density of the particles and $\sigma$ is their
diameter. Van Zon and Van Beijeren made an approximate calculation of the
largest Lyapunov exponent for a dilute gas of hard disks and obtained
results that are in very good agreement with the computer simulations of
Dellago\cite{vzvbd}. In addition van Beijeren, Dorfman, Posch, and Dellago
presented an approximate calculation of $h_{KS}/N$, the Kolmogorov-Sinai
entropy per particle, where $N >>1$ is the number of particles in the
system, for dilute gases of particles with short range forces in equilibrium%
\cite{vbdpd}. 
For hard sphere and hard disk systems their calculations led to expressions
for the KS entropy per particle of the form 
\begin{equation}
h_{KS}/N = a \nu [ -\ln \tilde{n} + b + O(\tilde{n})].  \label{1}
\end{equation}
Here $\nu$ is the average collision frequency per particle in the gas; $%
\tilde{n}=n\sigma^{2}$ for $d=2$, and $\tilde{n}=n\pi\sigma^{3}$ for $d=3$
is the reduced density of scatterers; and $a,b$ are numerical constants
depending on the number of spatial dimensions of the system. The values of $a
$ are in excellent agreement with computer simulations but the theoretical
values of $b$ differed from the computer results by a factor of about $3$
for $d=3$ and about $6$ for $d=2$.

The purpose of this paper is to provide a systematic method for computing $%
h_{KS}/N$, or in more general circumstances, the sum of all the positive
Lyapunov exponents, for a system of $N$ particles interacting with short
range forces. 
In principle, this method should give exact values for the coefficients $a$
and $b$ above, 
as well as higher density corrections to these results. 
It employs 
BBGKY hierarchy techniques 
for calculating the KS entropy per particle. 
Very similar hierarchy methods have been used in the kinetic theory of
moderately dense gases to derive the Boltzmann transport equation for $f(%
\vec{r}, \vec{v},t)$, and to generalize this equation to moderately dense
gases\cite{dvb}. Here $f$ is the one particle distribution function which
gives the probability of finding a particle in the gas at the spatial point $%
\vec{r}$, with velocity $\vec{v}$ at time $t$. Here we show that it is
possible to include new variables in the distribution function which
describe the separation of trajectory bundles in the phase space of $N$
particles. We can then extend and generalize the Boltzmann equation so as to
provide a method for computing $h_{KS}$ for dilute and moderately dense
gases in equilibrium and non-equilibrium steady states. In order to simplify
the discussion we will consider hard-sphere systems in $d$ dimensions and
briefly indicate at various points how these results can be extended to more
general potentials.

This paper is organized as follows: In Section II we describe the separation
of two trajectories in phase space in terms of a set of $d \times d$
matrices whose elements have the dimension of a time, and which we call the
radius of curvature (ROC) matrices. We then show that the sum of the
positive Lyapunov exponents can, for a hard sphere system in $d$ dimensions,
be written in terms of the ROC matrices and a reduced two-particle
distribution function. In Section III we describe the time dependence of
these matrices due to free motion of the particles and due to binary
collisions. After showing that the time dependence of the ROC matrices can
be expressed in terms of free streaming and binary collision operators, we
show that the reduced distribution functions defined in Section II satisfy a
set of hierarchy equations, similar to those appearing in the usual BBGKY
hierarchy equations, but modified to include the ROC matrices. In Section IV
we consider the low density case, and show that the sum of positive Lyapunov
exponents at low densities can be computed in terms of a single particle
distribution function which satisfies an extended Boltzmann transport
equation. In Section V we show how the previous results of 
Van Beijeren, Dorfman, Posch, and Dellago \cite{vbdpd} for the KS entropy
per particle, for an equilibrium system at low densities, can be obtained
from a partial solution of this extended Boltzmann equation for equilibrium
systems, and argue that an important contribution is neglected in this
approximate solution. Due to the complicated method that is 
needed to compute a better approximation to $h_{KS}/N$, we postpone the
calculation to a subsequent paper. We mention the new results in Section VI,
discuss some important issues and indicate some directions for further work.
Some of the most technical points are presented in the Appendices.

\section{The Separation of Phase Space Trajectories}

Generally, Lyapunov exponents characterize the rate at which infinitesimally
close trajectories in phase space separate or 
converge in the course of time\cite{ru-eck}. Here we consider a classical
system of $N$ identical particles, each of mass $m$, interacting with short
range forces in $d$ spatial dimensions. The phase space of such a system is $%
2dN$ dimensional and any trajectory of the system given by $\Gamma(t) = (%
\vec{r}_{1}(t), \vec{v}_{1}(t), \vec{r}_{2}(t), \vec{v}_{2}(t), ...,\vec{r}%
_{N}(t), \vec{v}_{N}(t))$ can be specified by giving the Hamiltonian of the
system and an initial point, $\Gamma(0)$. Here $\vec{r}_{i}(t),\vec{v}_{i}(t)
$ are the position and velocity of particle $i$ at time $t$, and we denote a
general phase point 
by $\Gamma$. To describe the Lyapunov exponents, we need to consider a
bundle of trajectories about $\Gamma(t)$ which are infinitesimally close to
it and which we denote by $\Gamma(t) + \delta\Gamma(t)$. The trajectory
deviations $\delta \Gamma(t)$ can be expressed in terms of the infinitesimal
spatial and velocity deviations for each particle as $\delta \Gamma =(\delta%
\vec{r}_{1}(t),\delta\vec{v}_{1}(t), ..., \delta \vec{r}_{N}(t), \delta \vec{%
v}_{N}(t))$.

To obtain an expression for the sum of the positive Lyapunov exponents we
recall some basic ideas of the chaotic dynamics of the phase space
trajectories of $N$ particle systems. One of the well-known consequences of
Liouville's theorem is that the Lebesgue measure of a set in the phase space 
$\Gamma$ is constant in time. This implies that if we pick a small set of
points $\delta\Gamma(0)$ about an initial phase point $\Gamma(0)$, then the
measure of this small set $\mu(\delta\Gamma)$ will be constant in time as
the points evolve according to Hamiltonian dynamics. 
As a consequence the sum of all of the Lyapunov exponents must be zero in
order that the measure be preserved in time. However, if we consider a
projection of $\delta\Gamma$ onto a subspace of lower dimension than the
full phase space, the measure of this projection in the reduced phase space
may change exponentially with time. Suppose then that we project $%
\delta\Gamma(0)$ onto the velocity subspace. Unless this projection is
accidentally pathological, its measure in velocity space will grow
exponentially in time with an exponential given by the sum of all the
positive Lyapunov exponents of the system 
(remember that for a time reversal invariant system the Lyapunov exponents
come in conjugate pairs $\lambda_i=-\tilde{\lambda}_i$, so the larger $dN$
Lyapunov exponents have to be $\ge 0$). If we denote the projection of $%
\delta\Gamma$ onto velocity space by $\delta{\cal V}$, and its volume in
velocity space by $\|\delta{\cal V}\|$, we expect that 
\begin{equation}
\|\delta{\cal V}(t)\| \sim \|\delta{\cal V}(0)\|\exp{[t
\,\sum_{\lambda_{i}>0}\lambda_{i}]},  \label{2}
\end{equation}
asymptotically for large $t$, where the Lyapunov exponents are denoted by $%
\lambda_{i}$. Before we turn to a method for computing the sum of positive
Lyapunov exponents, let us write Eq. (\ref{2}) in a form that will be useful
for our further work. That is, we write the sum of the positive Lyapunov
exponents formally as a time average. Then by making the assumption that the
system is ergodic\cite{kathas}, that is, its phase space consists of only
one ergodic component, we can replace the time average by an ensemble
average, as 
\begin{equation}
\sum_{\lambda_{i}>0}\lambda_{i}=\lim_{t\rightarrow \infty}\frac{1}{t}%
\ln\left[\frac{\|\delta{\cal V}(t)\|}{\|\delta{\cal V}(0)\|}\right]
=\lim_{t\rightarrow\infty}\frac{1}{t}\int_0^t d\tau \frac{d\ln\|\delta{\cal V%
}(\tau)\|}{d \tau} = \left<\frac{d\|\ln\delta{\cal V}(\tau)\|}{d \tau}%
\right>,  \label{3}
\end{equation}
where the angular brackets denote an ensemble average, over an appropriate
phase space ensemble, to be specified further on. If the limit on the left
hand side of Eq. (\ref{3}) exists and is non-zero, the system has positive
Lyapunov exponents and is chaotic. Under the above assumption about
ergodicity the ensemble average will provide an expression for the sum of
the positive Lyapunov exponents.

Now we imagine that the dynamics of the system can be characterized as a
sequence of instantaneous binary collisions connected by free motion of the
particles. We neglect all effects of collisions where three or more
particles are interacting at once, and all effects due to the finite
duration of the binary collisions. 
These assumptions are exact for hard core potentials but we are neglecting
small 
effects, which are of higher order in a density expansion, in the case that
the interaction potential is short range but not purely of the hard core
type. This can be seen by observing that the effects we neglect in such
cases are of the order of the duration of a collision divided by the mean
free time between collisions for a typical particle, or of the range of the
forces divided by the mean free path length. Between collisions the space
and velocity variables for all of the particles change with time as 
\begin{eqnarray}
\dot{\vec{r}}_{i}(t) = \vec{v}_{i}(t)  \nonumber \\
\delta \dot{\vec{r}}_{i}(t) = \delta\vec{v}_{i}(t)  \nonumber \\
\dot{\vec{v}}_{i}(t) = 0  \nonumber \\
\delta \dot{\vec{v}}_{i}(t) = 0,  \label{4}
\end{eqnarray}
where we assume, for the time being, that no external forces act on the
particles. Since all collisions are assumed to be instantaneous binary
collisions we can consider the changes in these variables at the instant of
one such collision between particles $k,l$ say. If we denote the variables
immediately after the $(k,l)$-binary collision by primes, then there are no
changes at the instant of the $k,l$ collision in the quantities $\vec{r}%
_{i}(t), \vec{v}_{i},\delta\vec{r}_{i}(t)$, and $\delta\vec{v}_{i}(t)$ for $%
i \neq k,l$, but for the colliding particles, the changes in these
quantities are given by 
\begin{eqnarray}
\vec{r}^{\prime}_{k} = \vec{r}_{k}; \,\, \vec{r}^{\prime}_{l} = \vec{r}_{l} 
\nonumber \\
\vec{V}^{\prime}_{kl} = \vec{V}_{kl}  \nonumber \\
\vec{v}^{\prime}_{kl} = {\bf U}_{kl}\cdot \vec{v}_{kl}  \nonumber \\
\delta\vec{R}^{\prime}_{kl} =\delta\vec{R}_{kl}; \,\,\,\delta\vec{V}%
^{\prime}_{kl} =\delta\vec{V}_{kl}  \nonumber \\
\delta\vec{r}^{\prime}_{kl} = {\bf U}_{kl}\cdot \delta\vec{r}_{kl}  \nonumber
\\
\delta\vec{v}^{\prime}_{kl} ={\bf U}_{kl}\cdot\delta\vec{v}_{kl} +{\bf \Omega%
}_{kl}\cdot\delta\vec{r}_{kl}.  \label{5}
\end{eqnarray}
Here $\vec{R}_{kl} = (1/2)(\vec{r}_{k}+\vec{r}_{l})$ is the location of the
center of mass of particles $k,l$ at the instant of collision, $\vec{V}%
_{kl}=(1/2)(\vec{v}_{k}+\vec{v}_{l})$ is the velocity of the center of mass
of the two particles, and $\vec{r}_{kl}=\vec{r}_{k}-\vec{r}_{l}, \vec{v}%
_{kl}=\vec{v}_{k}-\vec{v}_{l}$ are the relative position and velocity of
particles $k,l$. Furthermore, ${\bf U}_{kl}={\bf 1}-2\hat{\sigma}\hat{\sigma}
$ is a reflection operator about 
a plane orthogonal to the unit vector $\hat{\sigma}$, where $\hat{\sigma}$
is in the direction of the line connecting the center of particle $l$ to
that of $k$ at the point of closest approach of the two particles at their
collision and ${\bf 1}$ is the unit tensor in $d$ dimensions. For a
collision to take place we must have $(\vec{v}_{kl}\cdot\hat{\sigma})<0$.
The operator ${\bf \Omega}_{kl}$ appearing in the equation for the change in
the relative velocity deviation at collision is given by 
\begin{equation}
{\bf \Omega}_{kl}=\frac{2}{\sigma}\left[\vec{v}_{kl}\hat{\sigma} -\hat{\sigma%
}\vec{v}_{kl} -(\vec{v}_{kl}\cdot\hat{\sigma}){\bf 1} +\frac{\hat{\sigma}%
\hat{\sigma}\,v^{2}_{kl}}{(\vec{v}_{kl}\cdot\hat{\sigma})}\right].  \label{6}
\end{equation}
The above equations for the change in the relative velocity at collision are
well known, and recently Dellago, Posch, and Hoover have given a derivation
for the changes in the spatial and velocity deviations at collision, given
above, for hard sphere potentials\cite{delpoho}. It may look surprising at
first that $\delta\vec{r}^{\prime}_{kl}\neq \delta\vec{r}_{kl}$, because
positions do not change in a hard sphere collision. The difference is due,
however, to the infinitesimal difference in the time of the $k,l$ collision
on the two nearby trajectories. The results also apply to other short range
potentials provided one neglects {\it all} effects due to the finite
duration of a binary collision, including the motion of 
all of the particles during the $k,l$ collision, and provided one identifies 
$\hat{\sigma}$ with the unit vector along the direction of closest approach
between $k$ and $l$.

The spatial and velocity deviations are somewhat awkward for calculations
and a more convenient representation for the rate of separation of
trajectories in phase space is obtained by supposing that the spatial and
velocity deviations are related through a set of matrices which we will
refer to as radius of curvature, or ROC, matrices. A similar idea was used
by Sinai in his original formulation of an expression for the
Kolmogorov-Sinai entropy of a many particle hard sphere system in terms of
the ensemble average of the trace of the so-called second fundamental
operator which depends on the dynamics of the $N$ particles in the system%
\cite{sinai}. Here we define a set of matrices which are more convenient for
our purposes, but the central idea is certainly due to Sinai. An elementary
discussion of the second fundamental operator is given by Gaspard and Dorfman%
\cite{gasdo}. We suppose that the spatial deviation for particle $i$ at time 
$t$, $\delta\vec{r}_{i}(t)$, is related to the velocity deviations of all
the particles in the system through a set of matrices $\mbox{\boldmath
$\rho$}_{ij}$ by 
\begin{equation}
\delta\vec{r}_{i}(t)=\sum_{j}\mbox{\boldmath $\rho$}_{ij}(t)\cdot\delta\vec{v%
}_{j}(t)\,\,\,\,\,{\rm for}\,\, i=1,...N.  \label{7}
\end{equation}
While this relation couples the spatial deviation of one particle to the
velocity deviations of all of the particles, we will see below that this
coupling is quite manageable, and at low density is limited to the coupling
of the deviations for a colliding pair of particles. It is important to note
here that all of the expressions for the sum of the positive Lyapunov
exponents to be obtained below will depend only 
on the variables $\vec{r}_{i},\vec{v}_{i}$ for $i=1,...,N$, and on $%
\mbox{\boldmath $\rho$}_{ij}$ for $i,j = 1,...,N$ and not 
explicitly on the spatial and velocity deviations for the particles.

According to Eq. (\ref{3}) we need an expression for the time derivative of $%
\Vert \delta {\cal V}(t)\Vert $. Since there are no external forces acting
on the system, the volume of the velocity-space projection, $\delta {\cal V}$%
, remains constant whenever the particles are in free motion and only
changes with time due to binary collisions taking place in the gas. Let us
now consider the change in $\Vert \delta {\cal V}\Vert $ due to a binary
collision. The velocity deviations after a binary collision of particles $k$
and $l$ are related to those before collision by a matrix ${\bf A}_{kl}$ as 
\begin{equation}
\left( 
\begin{array}{c}
\delta \vec{v}_{1}^{\prime } \\ 
\delta \vec{v}_{2}^{\prime } \\ 
\vdots  \\ 
\delta \vec{v}_{N}^{\prime }
\end{array}
\right) ={\bf A}_{kl}\cdot \left( 
\begin{array}{c}
\delta \vec{v}_{1} \\ 
\delta \vec{v}_{2} \\ 
\vdots  \\ 
\delta \vec{v}_{N}
\end{array}
\right)   \label{8}
\end{equation}
where the primed variables denote the velocity deviations immediately after
the $(k,l)$ collision and the unprimed 
ones denote the velocity deviations immediately before collision. Of course,
only the velocity deviations for particles $k,l$ are affected by the
collision. Now the changes in the velocity deviations for particles $k,l$
are given by Eq. (\ref{5}), which involve the spatial deviations for these
two particles immediately before collision. These, in turn, are related to
the velocity deviations of all the particles through the ROC matrices
defined in Eq. (\ref{7}). Putting these equations together and noting that
the volumes in velocity space before and after the $(k,l)$ collision are  
related by 
\begin{equation}
\Vert \delta {\cal V}^{\prime }\Vert =|\det {\bf A}_{kl}|\,\Vert \delta 
{\cal V}\Vert ,  \label{9}
\end{equation}
we find that we will need the determinant of ${\bf A}_{kl}$ which is given
by 
\begin{equation}
\det {\bf A}_{kl}=\det \left[ {\bf U}_{kl}+\frac{1}{2}{\bf \Omega }%
_{kl}\cdot (\mbox{\boldmath $\rho$}_{kk}+\mbox{\boldmath $\rho$}_{ll}-%
\mbox{\boldmath $\rho$}_{kl}-\mbox{\boldmath $\rho$}_{lk})\right] .
\label{10}
\end{equation}
Besides Eqs. (\ref{5}) and (\ref{7}) we have used here the relation, $\delta 
\vec{V}_{kl}^{\prime }=\delta \vec{V}_{kl}$ as well as the fact that the
velocity deviations of all the other particles are unaffected by the $(k,l)$
collision. In Eq. (\ref{3}) above, we have expressed the sum of the positive
Lyapunov exponents in terms of the ensemble average of the time derivative
of $\ln \Vert \delta {\cal V}\Vert $. Since this quantity only changes at
the instant of a binary collision, we need to
({\bf a}) look at the binary
collisions taking place in the gas at some instant; ({\bf b}) calculate the
change in $\ln \Vert \delta {\cal V}\Vert $ at each of them; and ({\bf c})
take the average over an appropriate stationary ensemble distribution. Since
we have a total of $N$ particles in the system we need to count the possible
contributions from each of $N(N-1)/2$ possible binary collisions in the gas.
Thus we find that the sum of the positive Lyapunov exponents is given by  
\begin{equation}
\sum_{\lambda _{i}>0}\lambda _{i}=\sum_{k<l}\left\langle \sigma ^{d-1}\int d%
\hat{\sigma}|\vec{v}_{kl}\cdot \hat{\sigma}|\Theta (-\vec{v}_{kl}\cdot \hat{%
\sigma})\delta (\vec{r}_{kl}-\sigma \hat{\sigma})\ln |\det {\bf A}%
_{kl}|\right\rangle .  \label{11}
\end{equation}
Here the step function $\Theta (x)=1$ for $x>0$ and zero otherwise. We have
integrated over all possible values of $\hat{\sigma}$ for each collision,
included the fact that the centers of the particles must be separated by a
distance of $\sigma $ at collision, and included that rate at which
collisions take place in the factor $|\vec{v}_{kl}\cdot \hat{\sigma}|$ in
order to get the proper expression for the time derivative of the velocity
space volume. This expression is exact for hard-sphere systems in any number
of dimensions, and can be used, with appropriate modifications to the
differential scattering cross section, as a good approximation for other
short range potentials.

To proceed further we must explain how the ensemble average is to be
computed. We suppose that it is possible to find a stationary distribution
function which describes not only the space and velocity variables, $\vec{r}%
_{i},\vec{v}_{i}$, for each particle, but also the distribution of the $N^{2}
$ ROC matrices. While this seems especially complicated, it will turn out
that only some reduced one and two particle distribution functions are
actually needed for our computations. 
We assume the existence of a
distribution function ${\cal F}_{N}(x_{1},...,x_{N},\mbox{\boldmath $\rho$}%
_{11},\mbox{\boldmath $\rho$}_{12},...,\mbox{\boldmath $\rho$}_{NN})$, which
is normalized as 
\begin{equation}
\int dx_{1}...dx_{N}d\mbox{\boldmath $\rho$}_{11}...d\mbox{\boldmath $\rho$}%
_{NN}{\cal F}_{N}=1.  \label{12}
\end{equation}
here $x_{i}=(\vec{r}_{i},\vec{v}_{i})$ and the integrations are over all
components of position and velocity for each particle as well as over all
the elements of each of the ROC matrices. A simplification occurs
immediately since the ensemble average is over functions of two particles.
If we assume that ${\cal F}_{N}$ is symmetric in its variables we can
express the sum of the positive Lyapunov exponents in terms of a two
particle distribution function ${\cal F}_{2}(x_{1},x_{2},\mbox{\boldmath
$\rho$}_{11},\mbox{\boldmath $\rho$}_{12},\mbox{\boldmath $\rho$}_{21},%
\mbox{\boldmath $\rho$}_{22})$, as 
\begin{equation}
\sum_{\lambda _{i}>0}\lambda _{i}=\frac{\sigma ^{d-1}}{2}\int dx_{1}dx_{2}d%
\mbox{\boldmath $\rho$}_{11}...d\mbox{\boldmath $\rho$}_{22}\int d\hat{\sigma%
}|\vec{v}_{12}\cdot \hat{\sigma}|\Theta (-\vec{v}_{12}\cdot \hat{\sigma})%
{\cal F}_{2}\delta (\vec{r}_{12}-\sigma \hat{\sigma})\ln |\det {\bf A}_{12}|.
\label{13}
\end{equation}
Here the two particle distribution function is defined by 
\begin{equation}
{\cal F}_{2}(x_{1},x_{2},\mbox{\boldmath $\rho$}_{11},\mbox{\boldmath $\rho$}%
_{12},\mbox{\boldmath $\rho$}_{21},\mbox{\boldmath $\rho$}_{22})=\frac{N!}{%
(N-2)!}\int dx_{3}...dx_{N}\Pi ^{^{\prime }}\,d\mbox{\boldmath $\rho$}_{ij}%
{\cal F}_{N}(x_{1},...,\mbox{\boldmath $\rho$}_{NN}),  \label{14}
\end{equation}
where the prime on the product means that the elements of the matrices $%
\mbox{\boldmath $\rho$}_{11},\mbox{\boldmath $\rho$}_{12},\mbox{\boldmath
$\rho$}_{21},\mbox{\boldmath $\rho$}_{22}$ are not to be included
among the
integration variables. This is the main result of this section - an
expression for the sums of the positive Lyapunov exponents in terms of an
extended, stationary two particle distribution function. In the next section
we shall derive the BBGKY hierarchy equations that are needed to determine
this function.

\section{The BBGKY Hierarchy Equations}

For hard sphere systems one can describe the time evolution of any phase
space quantity in terms of a time evolution or ''pseudo-Liouville''
operator, ${\cal L}_{+}(N)$, such that any dynamical quantity $A(\Gamma (t))$
that does not have an explicit time dependence changes with time as \cite
{evlhd,hvb,de} 
\begin{equation}
A(\Gamma (t))=\exp [t\,{\cal L}_{+}]\,A(\Gamma (0)).  \label{15}
\end{equation}
Here the subscript $+$ indicates that we are considering the forward time
evolution, $t>0$, of the system, since the operators differ slightly for
evolution with $t<0$,
due to the different ways they  are defined for
nonphysical initial configurations. The phase point $\Gamma (t)$ is
physically meaningful only if no pair of particles is separated by a
distance less than $\sigma $, but even for unphysical phase points the time
evolution operator is well-defined. To be sure, for obtaining physically
meaningful results one has to multiply expressions of the form of the right
hand side of Eq, (\ref{15}) by a weighting function containing the factor $%
W(\Gamma )$ defined to be zero for overlapping configurations and unity for
non-overlapping configurations. 
Here we extend the definition of the pseudo-Liouville operator to include
the time dependence of the ROC matrices in addition to the positions and
velocities of all of the particles. We will then discuss the adjoint of this
operator which is used to obtain the BBGKY equations for the extended
distribution functions ${\cal F}_{n}$ for $n=1,2,...$. Since the method for
constructing the pseudo-Liouville operator has been discussed in the
literature at some length, and since the calculation here is an obvious
extension of this method, we merely outline the calculation here, following
the procedure in Ref. \cite{de}.

Suppose we had a system with only one particle. Then the time evolution of
some function $A(\vec{r}_{1},\vec{v}_{1},\mbox{\boldmath $\rho$}_{11})$ is
trivial: 
\begin{equation}
A(\vec{r}_{1},\vec{v}_{1},\mbox{\boldmath $\rho$}_{11},t) = A(\vec{r}_{1}+%
\vec{v}_{1}t,\vec{v}_{1}, \mbox{\boldmath $\rho$}_{11}+{\bf 1}t) =\exp[t%
{\cal L}_{0}(1)]A(\vec{r}_{1},\vec{v}_{1},\mbox{\boldmath $\rho$}_{11})
\label{16}
\end{equation}
where 
\begin{equation}
{\cal L}_{0}(1) = \vec{v}_{1}\cdot\frac{\partial}{\partial\vec{r}_{1}} +
\sum_{\alpha = 1}^{d}\frac{\partial}{\partial \mbox{\boldmath $\rho$}%
_{11,\alpha\alpha}}.  \label{17}
\end{equation}
Here we assume that there are no external forces acting on the system, the
derivatives are taken with respect to the spatial coordinates and with
respect to the {\it diagonal} elements of the ROC matrix. This latter
condition follows from the observation that under free particle motion, only
the diagonal parts of the ROC matrices change with time, growing linearly as
indicated above. Now consider the less trivial case of two particles. Let $%
A(x_{1},x_{2},\mbox{\boldmath $\rho$}_{11},\mbox{\boldmath $\rho$}_{12},%
\mbox{\boldmath $\rho$}_{21},\mbox{\boldmath $\rho$}_{22})$ be some function
of the indicated variables and consider the quantity ${\cal I}_{t}A$ given
by 
\begin{equation}
{\cal I}_{t}A = \left[{\cal S}_{t}(1,2)-{\cal S}_{t}^{0}(1,2)%
\right]A(x_{1},x_{2},\mbox{\boldmath $\rho$}_{11},...,\mbox{\boldmath $\rho$}%
_{22})  \label{18}
\end{equation}
where 
${\cal S}_{t}(1,2)$ replaces all of the dynamical variables by their values
at a time $t$ later, obtained by following the motion of particles $1$ and $2
$ forward in time, and ${\cal S}_{t}^{0}(1,2)$ is a free streaming operator
that acts on $A$ to produce 
\begin{eqnarray}
{\cal S}_{t}^{0}(1,2)A(\vec{r}_{1},\vec{v}_{1},\vec{r}_{2},\vec{v}_{2}, %
\mbox{\boldmath $\rho$}_{11},\mbox{\boldmath $\rho$}_{12},\mbox{\boldmath
$\rho$}_{21},\mbox{\boldmath $\rho$}_{22})  \nonumber \\
=A(\vec{r}_{1} + \vec{v}_{1}t,\vec{v}_{1},\vec{r}_{2}+\vec{v}_{2}t, \vec{v}%
_{2},\mbox{\boldmath $\rho$}_{11} + {\bf 1}t,\mbox{\boldmath $\rho$}_{12},%
\mbox{\boldmath $\rho$}_{21},\mbox{\boldmath $\rho$}_{22}+{\bf 1}t) 
\nonumber \\
=\exp[t{\cal L}_{0}(1,2)]A(\vec{r}_{1},...,\mbox{\boldmath $\rho$}_{22}).
\label{19}
\end{eqnarray}
where ${\cal L}_{0}(1,2)={\cal L}_{0}(1)+{\cal L}_{0}(2)$. By considering
the two particle dynamics and by using the method of Ref. \cite{de}, we can
express ${\cal I}_{t}A$ in terms of the free streaming operators and a
binary collision operator ${\cal T}_{+}^{(2)}(1,2)$ as 
\begin{equation}
{\cal I}_{t}A =\int_{0}^{t}d\tau{\cal S}^{0}_{\tau}(1,2) {\cal T}%
_{+}^{(2)}(1,2){\cal S}^{0}_{t-\tau}(1,2) A(\vec{r}_{1},...,\mbox{\boldmath
$\rho$}_{22}),  \label{20}
\end{equation}
where the the binary collision operator is given by 
\begin{equation}
{\cal T}_{+}^{(2)}(1,2)=\sigma^{d-1}\int d\hat{\sigma}|\vec{v}_{12}\cdot\hat{%
\sigma}|\Theta(-\vec{v}_{12}\cdot\hat{\sigma})\delta(\vec{r}_{12}-\sigma\hat{%
\sigma})({\cal P}_{\sigma}(1,2) -1).  \label{21}
\end{equation}
In Eq. (\ref{21}) ${\cal P}_{\sigma}(1,2)$ is a substitution operator which
replaces $\vec{v}_{1},\vec{v}_{2},\mbox{\boldmath $\rho$}_{11},%
\mbox{\boldmath $\rho$}_{12},\mbox{\boldmath $\rho$}_{21},\mbox{\boldmath
$\rho$}_{22}$ by their values immediately after a $(1,2)$ collision. These
values are given by Eq. (\ref{5}) as 
\begin{eqnarray}
\vec{v}^{\prime}_{1} = \vec{v}_{1} -(\vec{v}_{12}\cdot\hat{\sigma})\hat{%
\sigma}  \nonumber \\
\vec{v}^{\prime}_{2} =\vec{v}_{2} +(\vec{v}_{12}\cdot\hat{\sigma})\hat{\sigma%
}  \label{22}
\end{eqnarray}
for the velocities of the two colliding particles, and by the following
equations for the ROC matrices, derived in Appendix A, 
\begin{eqnarray}
\mbox{\boldmath $\rho$}^{\prime}_{a} +\mbox{\boldmath $\rho$}%
^{\prime}_{b}\cdot{\bf \Omega}_{12}\cdot\mbox{\boldmath $\rho$}_{d}=%
\mbox{\boldmath $\rho$}_{a},  \nonumber \\
\mbox{\boldmath $\rho$}^{\prime}_{b}\cdot{\bf U}_{12} +\mbox{\boldmath
$\rho$}^{\prime}_{b}\cdot{\bf \Omega}_{12}\cdot\mbox{\boldmath $\rho$}_{c}=%
\mbox{\boldmath $\rho$}_{b},  \nonumber \\
\mbox{\boldmath $\rho$}^{\prime}_{c}\cdot{\bf U}_{12} + \mbox{\boldmath
$\rho$}^{\prime}_{c}\cdot{\bf \Omega}_{12}\cdot\mbox{\boldmath $\rho$}_{c} =%
{\bf U}_{12}\cdot\mbox{\boldmath $\rho$}_{c},  \nonumber \\
\mbox{\boldmath $\rho$}^{\prime}_{d} +\mbox{\boldmath $\rho$}%
^{\prime}_{c}\cdot{\bf \Omega}_{12}\cdot\mbox{\boldmath $\rho$}_{d} = {\bf U}%
_{12}\cdot\mbox{\boldmath $\rho$}_{d},  \label{23}
\end{eqnarray}
where the new ROC matrices are linear combinations of the $\mbox{\boldmath
$\rho$}_{ij}$ given by 
\begin{eqnarray}
\mbox{\boldmath $\rho$}_{a} = \frac{1}{2}(\mbox{\boldmath $\rho$}_{11}+%
\mbox{\boldmath $\rho$}_{12}+\mbox{\boldmath $\rho$}_{21}+\mbox{\boldmath
$\rho$}_{22})  \nonumber \\
\mbox{\boldmath $\rho$}_{b} = \frac{1}{4}(\mbox{\boldmath $\rho$}_{11}-%
\mbox{\boldmath $\rho$}_{12}+\mbox{\boldmath $\rho$}_{21}-\mbox{\boldmath
$\rho$}_{22})  \nonumber \\
\mbox{\boldmath $\rho$}_{c} = \frac{1}{2}(\mbox{\boldmath $\rho$}_{11}-%
\mbox{\boldmath $\rho$}_{12}-\mbox{\boldmath $\rho$}_{21}+\mbox{\boldmath
$\rho$}_{22})  \nonumber \\
\mbox{\boldmath $\rho$}_{d} = (\mbox{\boldmath $\rho$}_{11}+\mbox{\boldmath
$\rho$}_{12}-\mbox{\boldmath $\rho$}_{21}-\mbox{\boldmath $\rho$}_{22}).
\label{24}
\end{eqnarray}
As usual, the primes denote values of the quantities immediately after a
collision.

The superscript $(2)$ on the binary collision operator denotes the condition
that we are only keeping two-particle contributions to the ROC matrix
expansions for $\delta \vec{r}_{1}$ and  $\delta \vec{r}_{2}$. In principle
we will have to keep many particle contributions to the ROC matrix
expansions of the spatial deviations which will force us to define a set of
binary collision operators, ${\cal T}_{+}^{(n)}(1,2|3,4,...,n)$, for $%
n=2,...,N$, even for a collision which only involves particles $1$ and $2$,
since the ROC matrices must be ``updated'' at each collision. While this
causes some complications in the formalism, these complications can be dealt
with along the lines used in the statistical mechanics of systems with more
than two body forces. Appendix B outlines the cluster expansion technique
needed to handle these extended binary collision operators, but for most of
the purposes of this paper, this cluster expansion will not be needed. We
mention here that unless there was some previous dynamical connection of
particle $k$, say, with either particle $1$ or $2$, the ROC matrices $%
\mbox{\boldmath $\rho$}_{1k},\mbox{\boldmath $\rho$}_{2k},\mbox{\boldmath
$\rho$}_{k1},\mbox{\boldmath $\rho$}_{k2}$ will all be zero before and after
the $(1,2)$ collision.

Now that we have defined a binary collision operator that acts on functions
of dynamical variables, we take its adjoint to get an operator that acts on
distribution functions. That is, consider an average value of ${\cal S}%
_{t}(1,2)A(x_{1},...,\mbox{\boldmath $\rho$}_{22})$ taken with respect to
some initial distribution function ${\cal F}_{2}(x_{1},...,\mbox{\boldmath
$\rho$}_{22},t=0)$ which we assume
contains a factor $W(1,2)$, the two
particle overlap function. The adjoint of the two-particle time displacement
operator is defined by 
\begin{eqnarray}
\lefteqn{\int dx_{1}dx_{2}d\mbox{\boldmath $\rho$}_{11}...d\mbox{\boldmath 
$\rho$}
_{22}{\cal F}_{2}(x_{1},...,\mbox{\boldmath $\rho$}_{22},t = 0){\cal S}
_{t}(1,2)A(x_{1},...,\mbox{\boldmath $\rho$}_{22})}  \nonumber \\
&=&\int dx_{1}dx_{2}d\mbox{\boldmath $\rho$}_{11}...d\mbox{\boldmath $\rho$}%
_{22}A(x_{1},x_{2},\mbox{\boldmath $\rho$}_{11},...,\mbox{\boldmath $\rho$}%
_{22})\bar{{\cal S}}_{-t}(1,2){\cal F}_{2}(x_{1},...,\mbox{\boldmath $\rho$}%
_{22},t=0),  \label{25}
\end{eqnarray}
where the ``bar'' denotes an adjoint, and the subscript notation on the
adjoint of the streaming operator calls attention to the fact that while
dynamical variables ``move forward'' in time, distribution functions `` move
backward'' in time.

After some elementary calculations one finds that there is a binary
collision operator representation of the adjoint streaming operator of the
form 
\begin{equation}
\bar{{\cal S}}_{-t}(1,2) = {\cal S}^{0}_{-t}(1,2) +\int^{t}_{0}d\tau {\cal S}%
^{0}_{-\tau}(1,2)\bar{{\cal T}}_{-}^{(2)}(1,2){\cal S}^{0}_{-(t-\tau)},
\label{26}
\end{equation}
where 
\begin{eqnarray}
\lefteqn{{\cal S}_{-t}^{0} = \exp[-t({\cal L}_{0}(1,2))] \,\,\, {\rm and}} \\
\lefteqn{\bar{{\cal T}}_{-}^{(2)}(1,2) = \sigma^{d-1}\int d\hat{\sigma}|\vec{%
v}_{12}\cdot\hat{\sigma}|\Theta(\vec{v}_{12} \cdot\hat{\sigma})\times} 
\nonumber \\
& & \left[\delta(\vec{r}_{12}-\sigma\hat{\sigma})\int d\mbox{\boldmath
$\rho$}^{\prime}_{11}d\mbox{\boldmath $\rho$}^{\prime}_{12}d\mbox{\boldmath
$\rho$}^{\prime}_{21} d{\mbox{\boldmath $\rho$}}^{\prime}_{22}%
\prod_{i,j =1,2}\delta(\mbox{\boldmath $\rho$}_{ij}-\mbox{\boldmath $\rho$}%
_{ij}(\mbox{\boldmath $\rho$}^{\prime}_{11},...,\mbox{\boldmath $\rho$}%
^{\prime}_{22})){\cal P}^{\prime}_{\sigma}(1,2) - \delta (\vec{r}_{12}+\sigma%
\hat{\sigma}) \right].  \label{27}
\end{eqnarray}
Here ${\cal P}^{\prime}_{\sigma}(1,2)$ is a substitution operator that
replaces ROC matrices to its right by the corresponding primed matrices, and
velocities by the restituting values, namely those which for a given $\hat{%
\sigma}$ produce $\vec{v}_{1},\vec{v}_{2}$ after a binary collision. The
delta functions in the ROC variables require that the primed ROC matrices be
the restituting ones, i.e., those which produce the unprimed ROC matrices
after a binary collision. These restituting ROC matrices are found by
interchanging the primed and non-primed matrices in Eq. (\ref{23}), and
solving for the primed matrices in terms of the unprimed ones.

The operator ${\cal T}_{+}^{(2)}(1,2)$ has the property that 
\begin{equation}
{\cal T}_{+}^{(2)}(1,2){\cal S}_{\tau }^{0}{\cal T}_{+}^{(2)}(1,2)=0
\label{29}
\end{equation}
since it is impossible for particles $1$ and $2$ to collide more than once
without the intervention of a third particle. A similar identity holds for
the $\bar{{\cal T}}_{-}^{(2)}(1,2)$ operator, so that we may write the time
displacement operators ${\cal S}_{t}(1,2)$ and $\bar{{\cal S}}_{-t}(1,2)$ in
a pseudo Liouville operator form 
\begin{equation}
{\cal S}_{t}(1,2)=\exp {t[{\cal L}_{0}(1,2)+{\cal T}_{+}^{(2)}(1,2)]},
\label{30}
\end{equation}
whenever the time displacement operator acts on dynamical variables and a
function $W(1,2)$ appears to the left of the time displacement operator, and 
\begin{equation}
\bar{{\cal S}}_{-t}(1,2)=\exp {-t[{\cal L}_{0}(1,2)-\bar{{\cal T}}%
_{-}^{(2)}(1,2)]},  \label{31}
\end{equation}
whenever the time displacement operator acts on distribution functions which
vanish for $r_{12}<\sigma $.

This analysis of the two particle time displacement operators can be readily
extended to the $N$ particle case, and it is this extension which allows us
to derive a set of BBGKY hierarchy equations for the $s$ particle
distribution functions defined by a natural generalization of Eq. (\ref{14}%
), 
\begin{equation}
{\cal F}_{s}(x_{1},...,x_{s},\mbox{\boldmath $\rho$}_{11},...,%
\mbox{\boldmath $\rho$}_{1s},\mbox{\boldmath $\rho$}_{21},...,%
\mbox{\boldmath $\rho$}_{ss},t) =\frac{N!}{(N-s)!}\int dx_{s+1}...dx_{N}{%
\prod}^\prime d\mbox{\boldmath $\rho$}_{ij}{\cal F}_{N}(x_{1},...,%
\mbox{\boldmath $\rho$}_{NN},t)  \label{32}
\end{equation}
where we have included a possible time dependence of the distribution
functions. The normalization used here is such that the successive
distribution functions, ${\cal F}_{s}$ are proportional to $(n\sigma^{d})^{s}
$, that is to the $s$-th power of the reduced density of the gas.

To derive the BBGKY equations we start with the $N$- particle
pseudo-Liouville operator and write the extended Liouville equation as 
\begin{equation}
\left[\frac{\partial}{\partial t}+{\cal L}_{0}(1,2,...,N) -\sum_{i<j}\bar{%
{\cal T}}_{-}^{(N)}(i,j|1,...,N)\right]{\cal F}_{N}=0,  \label{33}
\end{equation}
where ${\cal L}_{0}(1,2,...,N)=\sum_{i=1}^{N}{\cal L}_{0}(i)$ and the
N-particle binary collision operators are defined so as to include the
effects of the collision of particles $i$ and $j$ on all of the ROC
matrices. We hasten to add that unless there was some previous dynamical
connection of particle $k$, say, with either particle $i$ or $j$, the ROC
matrices $\mbox{\boldmath $\rho$}_{ik},\mbox{\boldmath $\rho$}_{jk},%
\mbox{\boldmath $\rho$}_{kj},\mbox{\boldmath $\rho$}_{ki}$ will all be zero
before and after the $(i,j)$ collision. This greatly simplifies the use of
these complicated looking binary collision operators. In Appendix B we show
that the $N$-particle binary collision operators can be expressed in terms
of a cluster expansion of the form 
\begin{equation}
\bar{{\cal T}}_{-}^{(N)}(1,2|3,...,N) =\bar{{\cal T}}_{-}^{(2)}(1,2)
+\sum_{n=3}^{N}\bar{{\cal U}}^{(n)}_{-}(1,2|3,...,n)  \label{34}
\end{equation}
where the $n$-particle cluster functions are defined recursively in Appendix
B. Here we give only the first one, 
\begin{equation}
\bar{{\cal U}}^{(3)}_{-}(1,2|3)=\bar{{\cal T}}_{-}^{(3)}(1,2|3)-\bar{{\cal T}%
}_{-}^{(2)}(1,2).  \label{35}
\end{equation}
If we now integrate over the pseudo-Liouville equation, Eq. (\ref{33}), and
use Eqs. (\ref{32}), and (\ref{35}), we obtain the BBGKY hierarchy
equations. In the following we will need only the first one 
\begin{eqnarray}
\lefteqn{\left[\frac{\partial}{\partial t}+{\cal L}_{0}(1)\right]
{\cal F}_{1}(x_1,
\mbox{\boldmath $\rho$}_{11}) =}  \nonumber \\
&&\int dx_{2}d\mbox{\boldmath $\rho$}_{12}d\mbox{\boldmath $\rho$}_{21}d
\mbox{\boldmath $\rho$}_{22}\bar{{\cal T}}^{(2)}_{-}(1,2){\cal F}%
_{2}(x_1,x_2,\mbox{\boldmath $\rho$}_{11},...\mbox{\boldmath $\rho$}_{22})
+\sum_{n=3}^{N}\int dx_{3}...d\mbox{\boldmath $\rho$}_{nn}\bar{{\cal U}}%
^{(n)}_{-}(1,2|3,...,n){\cal F}_{n}  \label{36}
\end{eqnarray}
We have not given the explicit form of the remaining hierarchy equations,
but they can easily be obtained, if so desired. When these equations are
integrated over all of the elements of the ROC matrices, then the usual
BBGKY equations are recovered. The extensions given here can be used for a
systematic determination of sums of Lyapunov exponents for dilute and
moderately dense gases. In addition they allow for non-systematic
approximations, similar to Enskog's equation for hard sphere systems \cite
{dvb,chapcow}, which may possibly be used even for systems in glassy states.

In the next section we truncate the hierarchy at the first equation to
obtain an extended Boltzmann equation for the sum of the positive Lyapunov
exponents at low density.

\section{The Extended Boltzmann Equation}

In the previous section we outlined the derivation of the full set of  the
extended BBGKY hierarchy equations. Their solution is a formidable task,
which has not been fulfilled even for the usual hierarchy equations.
Nevertheless substantial progress has been made toward approximately solving
the usual equations, leading to a more basic understanding and important
modification of the Enskog theory of dense hard sphere systems, to
explanations of long time tail phenomena in dense fluids, and to theories of
glasses, among other results \cite{dvb,trkd}. Here we will not attempt to
carry out a similar analysis for the extended equations - leaving that for
another place - but will restrict our attention to the theory for very low
densities.

We note from Eq. (\ref{13}) that the $2$-particle distribution function $%
{\cal F}_{2}$, is needed in order to determine the sum of the Lyapunov
exponents. Specifically, we need this function at the instant of a binary
collision of two particles $(1,2)$, say. From experience with the usual
hierarchy equations \cite{ed}, we propose that a useful expression for $%
{\cal F}_{2}$ at the instant just before collision in Eq. (\ref{13}) is
given by 
\begin{eqnarray}
{\cal F}_{2}(x_1,x_2,\mbox{\boldmath $\rho$}_{11},\mbox{\boldmath $\rho$}%
_{12},\mbox{\boldmath $\rho$}_{21},\mbox{\boldmath $\rho$}_{22},t)={\cal F}%
_{1}(x_1,\mbox{\boldmath $\rho$}_{11},t){\cal F}_{1}(x_2,\mbox{\boldmath
$\rho$}_{22},t)\delta(\mbox{\boldmath $\rho$}_{12})\delta(\mbox{\boldmath
$\rho$}_{21})  \nonumber \\
+{\cal G}_{2}(x_1,x_2,\mbox{\boldmath $\rho$}_{11},...,\mbox{\boldmath
$\rho$}_{22},t),  \label{38}
\end{eqnarray}
where we have introduced a new cluster function ${\cal G}_{2}$. In the first
term on the right hand side of Eq. (\ref{38}) we assume that particles $1$
and $2$ are totally uncorrelated at the instant of the collision. This not
only leads to the factorization of the pair distribution function, but also
to $\mbox{\boldmath $\rho$}_{12}=\mbox{\boldmath $\rho$}_{21}=0$, giving
rise to the delta functions in Eq. (\ref{38}). Any effects of a possible
correlation between the particles are put in ${\cal G}_{2}$. The Boltzmann
equation approximation, which we will use here, is to neglect the
contribution of the cluster function ${\cal G}_{2}$ in the expression for
the sum of the Lyapunov exponents. We then need to determine the
one-particle distribution function ${\cal F}_{1}$, which satisfies the first
hierarchy equation, Eq. (\ref{36}), where the right hand side depends upon
two and higher particle distribution functions. Again, we use the Boltzmann
approximation for ${\cal F}_{2}$, and neglect the contributions from the
higher order cluster operators and distribution functions on the right hand
side of Eq. (\ref{36}). These approximations can only be justified after one
has carefully examined the terms that are neglected, and shown that they are
of higher order in the density. We leave this for a further publication, and
explore here the effects of these approximations.

Under the above Boltzmann approximations we have an expression for the sum
of the positive Lyapunov exponents for low densities following from Eqs. (%
\ref{3}), (\ref{9}), (\ref{10}) as 
\begin{eqnarray}
\sum_{\lambda_{i}>0}\lambda_{i} \simeq \frac{\sigma^{d-1}}{2}\int dx_{1}
dx_{2}d\mbox{\boldmath $\rho$}_{11}d\mbox{\boldmath $\rho$}_{22}\int d\hat{%
\sigma}|\vec{v}_{12}\cdot\hat{\sigma}|\Theta(-\vec{v}_{12}\cdot\hat{\sigma}%
)\delta(\vec{r}_{12}-\sigma\hat{\sigma}) \times  \nonumber \\
\ln|\det\left[{\bf U}_{12}+\frac{1}{2}{\bf \Omega}_{12}\cdot(\mbox{\boldmath
$\rho$}_{11}+\mbox{\boldmath $\rho$}_{22})\right]|{\cal F}_{1}(x_1,%
\mbox{\boldmath $\rho$}_{11}){\cal F}_{1}(x_2,\mbox{\boldmath $\rho$}_{22}).
\label{39}
\end{eqnarray}
Here the one particle distribution functions obey, in this approximation, a
steady-state, extended Boltzmann equation of the form for $i=1,2$, 
\begin{equation}
{\cal L}_{0}(i){\cal F}_{1}(x_i,\mbox{\boldmath $\rho$}_{ii}) = \int dx_{3}d%
\mbox{\boldmath $\rho$}_{i3}d\mbox{\boldmath $\rho$}_{3i}d\mbox{\boldmath
$\rho$}_{33}\bar{{\cal T}}^{(2)}_{-}(i,3){\cal F}_{1}(x_i,\mbox{\boldmath
$\rho$}_{ii}){\cal F}_{1}(x_3,\mbox{\boldmath $\rho$}_{33})\delta(%
\mbox{\boldmath $\rho$}_{i3})\delta(\mbox{\boldmath $\rho$}_{3i}).
\label{40}
\end{equation}
This equation has to be supplemented by a boundary condition at $%
\mbox{\boldmath $\rho$}_{ii}=0$. For $\rho_{ii}<0$, ${\cal F}_{1}$ has to
vanish since semi-convex scatterers never produce negative radii of
curvature if those were not present initially. The boundary conditions at
infinity follow from the requirement that the distribution functions,
when integrated over the ROC variables, reduce to the usual space and
velocity distribution functions.
We use a steady state distribution function since we are interested in the
sum of the Lyapunov exponents in either equilibrium states, or
non-equilibrium steady states. Here we consider an equilibrium state, and
look for solutions of Eq. (\ref{40}) which, upon integration over the
elements of $\mbox{\boldmath $\rho$}_{ii}$ reduce to the usual
Maxwell-Boltzmann equilibrium distribution function. Once we have determined 
${\cal F}_{1}(x_i,\mbox{\boldmath $\rho$}_{ii})$ for $i=1,2$ by solving Eq. (%
\ref{40}), we can compute the sum of the positive Lyapunov exponents using
Eq. (\ref{39}).

\section{The Simple Collision Damping Approximation}

In our previous work on the Lyapunov spectrum for the random Lorentz gas at
low densities, we obtained an extended Lorentz-Boltzmann equation\cite
{lorpaps}. This equation could be solved under equilibrium and
non-equilibrium steady state conditions to provide expressions for the
Lyapunov spectrum for the Lorentz gas which agree very well with the
computer simulation studies of Posch and Dellago. There we found that the
dominant contribution to the Lyapunov spectrum comes from the collisional
damping of the appropriate distribution functions for the elements of the
ROC matrices, obtained either by solving the extended Lorentz-Boltzmann
equation for the distribution of the ROC matrix elements, or from a more
heuristic kinetic theory analysis which shows that the distribution of the
ROC matrix elements is closely related to the free path distribution for
particles in the gas. A similar free path analysis was applied to the sum of
positive Lyapunov exponents for a dilute gas in equilibrium. There 
it was
found that the analysis provides an expression for this sum which is of the
form given by Eq. (\ref{1}) with excellent agreement with simulations for
the coefficient $a$, but less than satisfactory agreement for $b$ in Eq. (%
\ref{1}). Here we show that the free path results can be easily recovered
from Eqs. (\ref{39}) and (\ref{40}), by making a simple approximation in the
expression for the binary collision operators appearing in Eq. (\ref{40}). A
better analysis of this equation leads to improved values for $b$, while
leaving the values of $a$ unchanged. This improved analysis is somewhat
lengthy and will presented elsewhere. Here, we show how the previous results
are recovered.

We begin by noticing that the expression, Eq. (28), for the binary collision
operator $\bar{{\cal T}}_{-}^{(2)}(i,j)$ consists of two parts, a
restituting term which replaces variables by their restituting values,
denoted by primed variables, and a direct part which accounts for the loss
of particles with the desired values of the variables due to collisions. As
the above remarks suggest, a simple approximation to the extended Boltzmann
equation, which incorporates the effect of collisional damping of the
distribution of the ROC matrix elements, may be obtained by 
assuming the restituting part of the binary collision operator on the right
hand side of Eq. (\ref{40}) is proportional to to $\delta (\mbox{\boldmath
$\rho$}_{11})$.
Such an approximation leads immediately to 
\begin{equation}
{\cal L}_{0}{\cal F}_{1}(x_{i},\mbox{\boldmath $\rho$}_{ii})\approx -\sigma
^{d-1}\int dx_{3}d\mbox{\boldmath $\rho$}_{33}d\hat{\sigma}|\vec{v}%
_{i3}\cdot \hat{\sigma}|\Theta (-\vec{v}_{i3}\cdot \hat{\sigma})\delta (\vec{%
r}_{i3}+\sigma \hat{\sigma}){\cal F}_{1}(x_{i},\mbox{\boldmath $\rho$}_{ii})%
{\cal F}_{1}(x_{3},\mbox{\boldmath $\rho$}_{33}).  \label{41}
\end{equation}
Now we use the fact that the system is assumed to be in equilibrium. Then we
can assume that ${\cal F}_{1}(x_{i},\mbox{\boldmath $\rho$}_{ii})$ does not
depend upon $\vec{r}_{i}$, and that when one integrates over the ROC
variables in the one particle distribution function, one obtains the
Maxwell-Boltzmann velocity distribution function. That is 
\begin{equation}
\int d\mbox{\boldmath $\rho$}_{33}{\cal F}_{1}(\vec{v}_{3},\mbox{\boldmath
$\rho$}_{33})=n\varphi (\vec{v}_{3}),  \label{42}
\end{equation}
where $n=N/V$ is the number density of the gas, and 
\begin{equation}
\varphi (\vec{v})=
\left( \frac{\beta m}{2\pi }\right) ^{d/2}\exp -\frac{%
\beta mv^{2}}{2},  \label{43}
\end{equation}
where $\beta =(k_{B}T)^{-1}$, with $T$ the thermodynamic temperature of the
gas, and $k_{B}$ is Boltzmann's constant. Thus we are led to a simple,
approximate, differential equation for ${\cal F}_{1}(x_{i},\mbox{\boldmath
$\rho$}_{ii})$ 
\begin{equation}
\sum_{\alpha =1}^{d}\frac{\partial }{\partial \rho _{ii,\alpha \alpha }}%
{\cal F}_{1}(x_{i},\mbox{\boldmath $\rho$}_{ii})\approx -\nu (\vec{v}_{i})%
{\cal F}_{(}x_{i},\mbox{\boldmath $\rho$}_{ii}),  \label{44}
\end{equation}
where $\nu (\vec{v}_{i})$ is the equilibrium, low density collision
frequency for a particle with velocity $\vec{v}_{i}$ given by 
\begin{equation}
\nu (\vec{v}_{i})=n\sigma ^{d-1}\int d\vec{v}_{3}\int d\hat{\sigma}|\vec{v}%
_{13}\cdot \hat{\sigma}|\Theta (-\vec{v}_{13}\cdot \hat{\sigma})\varphi (%
\vec{v}_{3}).  \label{45}
\end{equation}

 In order
to complete the calculation of the sum of the Lyapunov exponents, we return
to Eq. (\ref{39}) and examine the determinant in more detail. Since the
operator ${\bf U}_{12}$ is its own inverse and the magnitude of its
determinant is unity, it follows that 
\begin{eqnarray}
\left| \det[{\bf U}_{12} + {\bf \Omega}_{12}\cdot(\mbox{\boldmath $\rho$}%
_{11}+\mbox{\boldmath $\rho$}_{22})/2]\right| =\left| \det[{\bf 1}+{\bf %
\Gamma}_{12}\cdot (\mbox{\boldmath $\rho$}_{11}+\mbox{\boldmath $\rho$}%
_{22})/2]\right| \,\,\,{\rm where} \\
{\bf \Gamma}_{12}= \frac{2}{\sigma} \left[\vec{v}_{12}\hat{\sigma}+\hat{%
\sigma}\vec{v}_{12} -(\hat{\sigma}\cdot\vec{v}_{12}){\bf 1}-\frac{v_{12}^{2}%
}{(\hat{\sigma}\cdot\vec{v}_{12})}\hat{\sigma}\hat{\sigma}\right].
\label{47}
\end{eqnarray}
It is easily seen that the operator ${\bf \Gamma}_{12}$ is proportional to a
projection operator onto a subspace orthogonal to the relative velocity
vector $\vec{v}_{12}$. We consider the two and three dimensional cases
separately since a bit of geometry is required now to complete the
calculation.

We begin with the case of two dimensions. Then the determinant we need is
found to be 
\begin{eqnarray}
\det [{\bf 1}+{\bf \Gamma }_{12}\cdot (\mbox{\boldmath $\rho$}_{11}+%
\mbox{\boldmath $\rho$}_{22})/2] &=&1+\frac{2|\vec{v}_{12}|}{\sigma |\cos
\phi |}\rho _{\perp \perp }\,\,\,{\rm where} 
\label{48}\\
\vec{v}_{12}\cdot \hat{\sigma} &=&|\vec{v}_{12}|\cos \phi ;\,\,{\rm and}%
\,\,\,\rho _{\perp \perp }=\frac{1}{2}(\hat{v}_{12\perp }\cdot (%
\mbox{\boldmath $\rho$}_{11}+\mbox{\boldmath $\rho$}_{22})\cdot \hat{v}%
_{12\perp }.  \label{49}
\end{eqnarray}
Here the subscript $\perp $ on a two dimensional unit vector $\hat{v}%
_{12,\perp }$ denotes a unit vector orthogonal $\hat{v}_{12}$. From Eq. (\ref
{49}) we see that we need the distribution for only one of the diagonal
components of the ROC matrices $\mbox{\boldmath $\rho$}_{11}$ and $%
\mbox{\boldmath $\rho$}_{22}$ in a representation based on the coordinate
frame defined by $\hat{v}_{12},\hat{v}_{12\perp }$. From Eq. (\ref{44}),
integrated over the other component, this distribution function, properly
normalized, is easily found to be 
\begin{equation}
{\cal F}_{1}(\vec{v}_{i},\rho _{ii,\perp \perp })=n\nu (\vec{v}_{i})\varphi (%
\vec{v}_{i})\exp [-\nu (\vec{v}_{i})\rho _{ii,\perp \perp }].  \label{50}
\end{equation}
When one uses this expression in Eq. (\ref{39}) and calculates the quantity $%
\sum_{\lambda _{i}>0}\lambda _{i}/N$,the KS entropy per particle for a
dilute gas of hard disks, one recovers exactly the expression obtained in
Ref. \cite{vbdpd}, given in Eq. (\ref{1}) above. We note that the first term
on the right hand side of Eq. (\ref{48}), can be neglected here since the
next term is on the order of the free time between collisions divided by the
time needed to travel a molecular diameter which is of order $\sigma /|\vec{v%
}_{12}|$. For a dilute gas this ratio is much larger than $1$ and the error
made in neglecting the first term is higher order in the density.

The three dimensional case is treated in a very similar way. In this case
the space orthogonal to $\vec{v}_{12}$ is two dimensional, and the
corresponding operator ${\bf \Gamma }_{12}$ projects the matrices $%
\mbox{\boldmath $\rho$}_{11},\mbox{\boldmath $\rho$}_{22}$ onto this
subspace. Thus, instead of needing the nine components of each of these
matrices, we need only four. We then find, for three dimensions 
\begin{equation}
\det \left[ {\bf 1}+{\bf \Gamma }_{12}\cdot \frac{1}{2}(\mbox{\boldmath
$\rho$}_{11}+\mbox{\boldmath $\rho$}_{22})\right] =\left( \frac{2|v_{12}|}{%
\sigma }\right) ^{2}[\rho _{\perp \perp }^{11}\rho _{\perp \perp }^{22}-\rho
_{\perp \perp }^{12}\rho _{\perp \perp }^{21}]  \label{51}
\end{equation}
where 
\begin{equation}
\rho _{\perp \perp }^{ij}={\hat{v}}_{12,\perp }^{(i)}\cdot \frac{(%
\mbox{\boldmath $\rho$}_{11}+\mbox{\boldmath $\rho$}_{22})}{2}\cdot {\hat{v}}%
_{12,\perp }^{(j)}
\equiv \frac{\rho _{11\bot \bot }^{ij}+\rho _{22\bot \bot
}^{ij}}{2},  \label{52}
\end{equation}
and the unit vectors $\hat{v}_{12},\hat{v}_{12.\perp }^{(1)},\hat{v}_{12,\perp }^{(2)}$
form an orthogonal coordinate basis in three dimensions. Of the four matrix
elements needed for each of the ROC matrices only the two diagonal
components (in superscript) are important here. This follows from the
observation that, according to Eq. (\ref{17}), between collisions the
diagonal components grow linearly with time, while the off-diagonal elements
remain constant, changing only at collisions. Since the time between
collisions scales inversely with the density, only the diagonal components
are important at low density for obtaining the dominant contribution to the
sum of the Lyapunov exponents. We then find that 
\begin{eqnarray}
h_{KS}^{(3)}/N &=&\frac{\sigma ^{2}}{n}\int d\vec{v}_{1}d\vec{v}_{2}d\rho
_{11,\perp \perp }^{11}d\rho _{22,\perp \perp }^{11}d\hat{\sigma}|\vec{v}%
_{12}\cdot \hat{\sigma}|\Theta (-\vec{v}_{12}\cdot \hat{\sigma})\ln \left[ 
\frac{|v_{12}|(\rho _{11,\perp \perp }^{11}+\rho _{22,\perp \perp }^{11})}{%
\sigma }\right]   \nonumber \\
&&{\cal F}_{1}(\vec{v}_{1},\rho _{11,\perp \perp }^{11}){\cal F}_{1}(\vec{v}%
_{2},\rho _{22,\perp \perp }^{11}).  \label{53}
\end{eqnarray}
The single particle distribution functions appearing in Eq. (\ref{53}) have
the same form as those in Eq. (\ref{50}) with the three dimensional form of
the collision frequency used in place of the two dimensional form. This too
is the same result as that obtained in Ref. \cite{vbdpd}

\section{Conclusion}

We have outlined here a method that allows a systematic expansion of the sum
of the positive Lyapunov exponents for a gas as a function of the gas
density. For a gas in equilibrium, this leads to an expression for the KS
entropy which, until recently\cite{vbdpd}, had been the subject of much
speculation but with no explicit analytic results\cite{gaswang}. We have
formulated the theory for a gas of hard spheres but it can be extended to
gases which interact with other short range potentials by extending the
definition of the binary collision operators to such potentials. Such an
extension will almost certainly ignore effects that take place on the time
scale of the duration of a binary collision as well as effects of genuine
many-particle collisions. Nevertheless, such an extension would provide
useful, if not completely accurate results. A similar situation obtains in
the kinetic theory for transport coefficients and time correlation functions
where hard sphere results can be extended to other potentials with similar
approximations\cite{dvb}. A simple way to extend the low density results to
higher densities, for hard sphere systems and other systems with short range
potentials, is to use the Enskog approximation when computing the
two-particle distribution function\cite{dvb,chapcow,vbe}. In this
approximation one replaces the two particle function at the instant of a
collision by the product of single particle functions as done here, but
multiplies this product by the pair distribution function for two hard
spheres at contact, the so-called Enskog $\chi$-factor. This approximation
leads to useful approximations for the transport coefficients and time
correlation functions of dense gases \cite{dvb,jrdedgc}.

There are now several clear directions for future work:

1) The low density results as derived here must be improved by avoiding the
approximation that the restituting collisions are proportional to $\delta(%
\mbox{\boldmath $\rho$}_{ii})$, 
as indicated in Section IV above. Some progress has been made in this
direction. One of us (JRD)
found, in an approximate calculation of the restituting contributions, that
the coefficient $a$ in Eq. (\ref{1}) is unchanged as expected, but there is
an addition to $b$ which is about $\ln4$ in two dimensions, changing $b$
from $0.2$ to $\approx 1.6$, compared to the computer result of $b\simeq 1.35
$, In three dimensions, an additional contribution to $b$ of approximately $%
\ln3$ is found, changing the previous result from $b=0.56$ to $b \approx 1.66
$, compared to the computer result of $b\simeq 1.34$ \cite{jrd}. These
results are now much closer to the computer results and an encouraging sign
of the correctness of this approach. It is worth remarking here that the
important role of the restituting collisions is due to the
``semi-dispersing" nature of hard sphere systems, which is quite different
from the Lorentz gas system which is a purely dispersing system. In a
collision between particles $k$ and $l$ not all of the components of the ROC
matrices $\mbox{\boldmath $\rho$}_{kk}$ 
and $\mbox{\boldmath $\rho$}_{ll}$ will be reduced to values on the order of 
$\sigma (m/k_{B}T)^{1/2})$, as is the case for the Lorentz gas. This will be
explained in a further publication \cite{dlvb2}.

2) These methods and results must be extended to other situations where sums
of Lyapunov exponents play an important role. Such situations include
trajectories on the fractal repeller in an open system, 
as discussed by Dorfman and Gaspard \cite{dogas}. In such systems the sum of
the positive Lyapunov exponents together with the KS entropy (which does not
satisfy Pesin's theorem in this case) determines the transport coefficients
of the system. Other situations of interest occur in systems subject to
external driving forces in combination with Gaussian thermostats. In such
systems 
the Liouville measure is not preserved. There is an interesting connection
between the sums of all the Lyapunov exponents and transport phenomena such
as entropy production and transport coefficients for such systems, 
which has
been discussed in detail in the literature\cite{thermostuff}. The methods
discussed here can be extended to treat such thermostatted systems, also,
but one has to calculate both the sums of all the positive and all the
negative Lyapunov exponents.

3) The method used here can be extended to higher densities by including the
successively higher hierarchy equations and correlation functions. In
addition one can make non-systematic high density approximations in the
spirit of the Enskog equation. Among other reasons it is important to look
at higher density effects since one can then make some connections between
dynamical systems theory and long time memory effects in a fluid. This will
be important not only for a deeper understanding of non-equilibrium
phenomena in general, but also for answering some of the questions that
arise in the theory of glasses, particularly those associated with the
ergodic properties of such systems \cite{glass}.

ACKNOWLEDGMENTS: We wish to thank Prof. M. H. Ernst and Mr. Ramses van Zon
for many helpful conversations, and the National Science Foundation for
support under grant PHY-96-00428. HvB acknowledges support by FOM, SMC, and
by the NWO Priority Program Non-Linear Systems, which are financially
supported by the ``Nederlandse Organisatie voor Wetenschappelijk Onderzoek
(NWO)". A.L. thanks the Deutsche Forschungsgemeinschaft for financial support
via the SFB 262. 
\appendix

\section{Derivation of Eq. (\ref{24})}

We begin by noting that for a system of two particles, the spatial deviation
vectors are expressed in terms of ROC matrices as 
\begin{eqnarray}
\delta \vec{r}_{1} &=&\mbox{\boldmath $\rho$}_{11}\cdot \delta \vec{v}_{11}+%
\mbox{\boldmath $\rho$}_{12}\cdot \delta \vec{v}_{2},  \nonumber \\
\delta \vec{r}_{2} &=&\mbox{\boldmath $\rho$}_{21}\cdot \delta \vec{v}_{11}+%
\mbox{\boldmath $\rho$}_{22}\cdot \delta \vec{v}_{2}.  \label{a1}
\end{eqnarray}
{F}rom this it follows that 
\begin{eqnarray}
\delta \vec{R}_{12} &=&\mbox{\boldmath $\rho$}_{a}\cdot \delta \vec{V}_{12}+%
\mbox{\boldmath $\rho$}_{b}\cdot \delta \vec{v}_{12},  \nonumber \\
\delta \vec{r}_{12} &=&\mbox{\boldmath $\rho$}_{c}\cdot \delta \vec{v}_{12}+%
\mbox{\boldmath $\rho$}_{d}\cdot \delta \vec{V}_{12},  \label{a2}
\end{eqnarray}
where $\mbox{\boldmath $\rho$}_{a},...,\mbox{\boldmath $\rho$}_{d}$ are
defined in Eq. (\ref{24}). Now consider the changes in $\delta \vec{R}_{12}$
and $\delta \vec{r}_{12}$ at collision as given by Eq. (5). The center of
mass equation becomes - the primed variable denote values after collision - 
\begin{equation}
\mbox{\boldmath $\rho$}_{a}^{\prime }\cdot \delta \vec{V}_{12}^{\prime }+%
\mbox{\boldmath $\rho$}_{b}^{\prime }\delta \vec{v}_{12}^{\prime }=%
\mbox{\boldmath $\rho$}_{a}\cdot \delta \vec{V}_{12}+\mbox{\boldmath $\rho$}%
_{b}\cdot \delta \vec{v}_{12}.  \label{a3}
\end{equation}
We now insert the expressions $\delta \vec{V}_{12}^{\prime }=\delta \vec{V}%
_{12}$ and and the appropriate one for $\delta \vec{v}_{12}^{\prime }$ given
by 
\begin{equation}
\delta \vec{v}_{12}^{\prime }={\bf U}_{12}\cdot \vec{v}_{12}+{\bf \Omega }%
_{12}\cdot [\mbox{\boldmath $\rho$}_{c}\cdot \delta \vec{v}_{12}+%
\mbox{\boldmath $\rho$}_{d}\cdot \delta \vec{V}_{12}],  \label{a4}
\end{equation}
into Eq. (\ref{a3}), to obtain 
\begin{equation}
\mbox{\boldmath $\rho$}_{a}^{\prime }\cdot \delta \vec{V}_{12}+%
\mbox{\boldmath $\rho$}_{b}^{\prime }\cdot \left[ {\bf U}_{12}\cdot \delta 
\vec{v}_{12}+{\bf \Omega }_{12}\cdot (\mbox{\boldmath $\rho$}_{c}\cdot
\delta \vec{v}_{12}+\mbox{\boldmath $\rho$}_{d}\cdot \delta \vec{V}%
_{12})\right] =\mbox{\boldmath $\rho$}_{a}\cdot \delta \vec{V}_{12}+%
\mbox{\boldmath $\rho$}_{b}\cdot \delta \vec{v}_{12}.  \label{a5}
\end{equation}
We can now equate coefficients of $\delta \vec{V}_{12}$ and of $\delta \vec{v%
}_{12}$ on each side of Eq. (\ref{a3}) since the relative and center of mass
velocities are independent of each other, and there are no orthogonality or
normalization contraints on the components of each of these velocity
deviation vectors which would make some of the components of each of these
vectors dependent on the others. One therefore finds the first two lines of
Eq. (\ref{23}). A similar treatment of the collision equation for $\delta 
\vec{r}_{12}$ leads to the third and fourth lines of Eq. (\ref{23}).

\section{Cluster 
Expansions of the Binary Collision Operators}

Cluster expansion methods have proven to be of great value in both the
equilibrium and the non-equilibrium statistical mechanics of fluid systems 
\cite{dvb,cohen,uhlford,jrdegdcdiv}. They allow us to express functions
defined for a system of $N$ particles in terms of functions defined for
systems of fewer numbers of particles. In equilibrium fluids, for example,
one can obtain the virial expansions of 
thermodynamic functions using Ursell
cluster expansions of the $N$-particle Boltzmann factor $\exp [-\beta H_{N}]$%
. Here we apply the Ursell method to the $N$ particle binary collision
operator $\bar{{\cal T}}_{-}^{(N)}(1,2|3,4,...,N)$. The idea is to choose
the first term in the expansion as a two particle function, with the next
and higher order terms successively correcting for the errors made in the
earlier terms. For example, for three particles, we would write 
\begin{equation}
\bar{{\cal T}}_{-}^{(3)}(1,2|3)=\bar{{\cal T}}_{-}^{(2)}(1,2)+\bar{{\cal U}}%
_{-}^{(3)}(1,2|3),  \label{b1}
\end{equation}
where, obviously, 
\begin{equation}
\bar{{\cal U}}_{-}^{(3)}(1,2|3)=\bar{{\cal T}}_{-}^{(3)}(1,2|3)-\bar{{\cal T}%
}_{-}^{(2)}(1,2).  \label{b2}
\end{equation}
This procedure can be repeated for four particles to define a four particle
cluster operator $\bar{{\cal U}}_{-}^{(4)}(1,2|3,4)$ through the equation 
\begin{equation}
\bar{{\cal T}}_{-}^{(4)}(1,2|3,4)=\bar{{\cal T}}_{-}^{(2)}(1,2)+\bar{{\cal U}%
}_{-}^{(3)}(1,2|3)+\bar{{\cal U}}_{-}^{(3)}(1,2|4)+\bar{{\cal U}}%
_{-}^{(4)}(1,2|3,4).  \label{b3}
\end{equation}
Extending this idea to $N$ particles, we find 
\begin{equation}
\bar{{\cal T}}_{-}^{(N)}(1,2|3,4,...,N)=\bar{{\cal T}}_{-}^{(2)}(1,2)+%
\sum_{k=3}^{N}\bar{{\cal U}}_{-}^{(3)}(1,2|k)+\sum_{k<l}\bar{{\cal U}}%
_{-}^{(4)}(1,2|k,l)+\cdots .  \label{b4}
\end{equation}
This cluster expansion of the binary collision operator is used to obtain
Eq. (\ref{36}). Similar cluster expansions. with simple modifications, can
be used for the $N$ particle distribution functions and pseudo Liouville
operators.

\end{document}